\documentclass[fleqn,usenatbib]{mnras}

\usepackage{newtxtext,newtxmath}

\usepackage[T1]{fontenc}
\usepackage{color}

\DeclareRobustCommand{\VAN}[3]{#2}
\let\VANthebibliography\thebibliography
\def\thebibliography{\DeclareRobustCommand{\VAN}[3]{##3}\VANthebibliography}

\newcommand{\obj}{J1606+3124}

\usepackage{graphicx}	
\usepackage{amsmath}	

\title[A CSO at $z=4.56$]{A compact symmetric radio source born at one-tenth the current age of the Universe}

\author[]{
Tao An $^{1,3,4}$\thanks{E-mail: antao@shao.ac.cn},
Ailing Wang $^{1,2}$\thanks{E-mail: wal@shao.ac.cn}, 
Yingkang Zhang$^{1}$,
J.N.H.S. Aditya$^{1}$,
Xiaoyu Hong$^{1,2}$, \newauthor
Lang Cui$^{4}$
\\
$^{1}$Shanghai Astronomical Observatory, Chinese Academy of Sciences, Nandan Road 80, Shanghai 200030, China \\
$^{2}$College of Astronomy and Space Sciences, University of Chinese Academy of Sciences, 19A Yuquanlu, Beijing 100049, China\\
$^{3}$ Key Laboratory of Cognitive Radio and Information Processing, Guilin University of Electronic Technology, 541004 Guilin, China \\
$^{4}$ Xinjiang Astronomical Observatory, Key Laboratory of Radio Astronomy, Chinese Academy of Sciences, 150 Science 1-Street, Urumqi 830011, China
}

\date{Accepted XXX. Received YYY; in original form ZZZ}

\pubyear{2020}

\begin{document}
\label{firstpage}
\pagerange{\pageref{firstpage}--\pageref{lastpage}}
\maketitle

\begin{abstract} 
Studies of high redshift radio galaxies can shed light on the activity of active galactic nuclei (AGN) in massive elliptical galaxies, and on the assembly and evolution of galaxy clusters in the Universe. 
\obj\ has been tentatively identified as a radio galaxy at a redshift of 4.56, at an era of one-tenth of the current age of the Universe. Very long baseline interferometry (VLBI) images show a compact triple structure with a size of 68 parsecs.
The radio properties of \obj, including the edge-brightening morphology, peaked GHz radio spectrum, slow variability, and low jet speed, consistently indicate that it is a compact symmetric object (CSO). The radio source size and expansion rate of the hotspots suggest that \obj\ is a young (kinematic age of $\sim$3600 years) radio source.  
Infrared observations reveal a gas- and dust-rich host galaxy environment, which may hinder the growth of the jet; however, the ultra-high jet power of \obj\ gives it  an excellent chance to grow into a large-scale double-lobe radio galaxy.
If its redshift and galaxy classification can be confirmed by further optical spectroscopic observations, \obj\ will be the highest redshift CSO galaxy known to date.
\end{abstract}




\section{Introduction}
\label{sec:intro}

High redshift radio galaxies (HzRGs) are the progenitors of the dominant elliptical galaxies in galaxy clusters in the local Universe, and clusters are the building blocks of the Universe \citep{2008A&ARv..15...67M}. The study of HzRGs sheds light on the assembly and evolution of large-scale structures in the Universe. Moreover, HzRGs are also among the most massive galaxies at their redshifts \citep{2009ApJ...704..548O}, and therefore fascinating objects on their own to study the activity of the active galactic nuclei (AGN) and the accretion of supermassive black holes (SMBHs). However, the challenges of observing HzRGs are significant, resulting in only a few have been observed at $z \gtrsim 4.5$, when the Universe was about one-tenth its current age \citep{2018MNRAS.480.2733S}. Despite the observational difficulties, an in-depth study of the few individual HzRGs that have been identified and confirmed is well worthwhile and necessary due to the critical value of the HzRGs in astrophysical research.

The source FIRST  J160608.5+312446 (also named as CRATES \obj; in short, \obj) is identified as a high-redshift object at $z=4.56$ \citep{2008ApJS..175...97H}\footnote{Due to its faintness, its spectroscopic measurement may have systematic errors \citep{2008ApJS..175...97H}. Until a new redshift is determined based on more sensitive optical spectroscopic observations, $z = 4.56$  is used in this paper.}. It is very faint in the optical band and showed an empty field \citep{1991ApJ...380...66O}. Later, \citet{1993ApJS...88....1S} obtained a CCD image of \obj, showing a faint and complex feature \citep{1993ApJS...88....1S}.  But it is bright in the infrared \citep[named WISEA J160608.53+312446.5,][]{2013wise.rept....1C} and radio bands \citep[e.g.,][]{1977AJ.....82..776O}. 
\obj\ has been detected as a bright source in several follow-up radio surveys, including the Green Bank 91 m telescope survey at 4.85 GHz  \citep{1991ApJS...75.1011G,1991ApJS...75....1B}, the Faint Images of the Radio Sky at Twenty Centimeters \citep[FIRST,][]{1995ApJ...450..559B} of the Very Large Array (VLA) and  NRAO VLA Sky Survey at 1.4 GHz \citep[NVSS,][]{1998AJ....115.1693C}, the VLBA Calibrator Survey \citep[VCS,][]{2002ApJS..141...13B} and the Combined Radio All-Sky Targeted Eight GHz Survey of the VLA at 8.4 GHz \citep[CRATES,][]{2007ApJS..171...61H}. 
It has also been detected in the ongoing surveys, such as the Very Large Array Sky Survey at 3 GHz \citep[VLASS,][]{2020RNAAS...4..175G} and the Australian Square Kilometre Array Pathfinder (ASKAP) continuum survey at 0.888 GHz \citep[the first large-area survey named RACS, ][]{2020PASA...37...48M}.

The radio source nature of \obj\ is a matter of debate. In early observational studies three decades ago, this source was identified as a gigahertz-peaked spectrum (GPS)  galaxy  \citep[e.g.,][]{1991ApJ...380...66O,1993ApJS...88....1S}. However, some other subsequent studies identified it as a flat-spectrum radio quasar \citep{2008ApJS..175...97H}. The reason for this divergence in radio classification stems from differences in the observation frequency coverage and possible variability in non-simultaneous observations. Recent simultaneous multi-frequency observations from the RATAN-600 radio telescope (in short, RATAN-600), covering a frequency range from 1.1 to 21.7 GHz, showed a convex spectrum with a turnover between 2 and 4 GHz \citep{2012A&A...544A..25M,2019AstBu..74..348S}, thus supporting it as a GPS source candidate. 

GPS sources are classified according to their radio spectral shape, and they are a hybrid class of quasars and galaxies, which are physically different sources. The redshift distributions of these two classes are distinctively different: GPS quasars are found at a wider redshift range ($1 \leq z \leq 4$) \citep{1991ApJ...380...66O}, while GPS galaxies tend to be found at relatively lower redshifts ($0.1 \leq z \leq 1$) with only a few at $z>2$ \citep{1996ApJ...470..806O,2012A&A...544A..25M}. This difference in redshift distribution is probably due to observational effects, \textit{i.e.}, quasars are more easily observed at higher redshifts than galaxies, in addition to other possible intrinsic mechanisms.
According to the AGN unified scheme \citep{1995PASP..107..803U}, the  difference between quasars and radio galaxies is caused by their different jet viewing angles: radio galaxies generally have larger jet viewing angles than quasars. Measuring the jet viewing angle requires high-resolution images of the jet. The published 5-GHz very long baseline interferometry (VLBI) image of \obj\ \citep{2007ApJ...658..203H} showed two compact components along the north-south direction on milli-arcsec (mas) scales, resembling a compact symmetric object (CSO).  CSOs are often thought of as powerful radio AGN in their early evolutionary stages \citep[e.g.,][]{2012ApJ...760...77A,2021A&ARv..29....3O}. 
\obj\ is at a cosmic era shortly after the end of reionisation, while AGN activity peaks at $z = 2 - 3$ \citep{1995AJ....110...68S,2000MNRAS.311..576K}. If \obj\ is indeed a CSO, it would be the most distant CSO galaxy known to date; then it would be an important target for studying how AGN activity is triggered during the important transitional phase towards its peak of activity. 

In addition to the conventional criteria for CSOs ($<$1 kpc radio source size, symmetric jet structure with respect to the core, GHz peaked spectrum shape), two additional criteria for CSOs have recently  been proposed by \citet{2021arXiv211108818R}: slow radio variability and low apparent jet advancing speed. 
To further clarify the radio classification of \obj\ and to study its radio properties, we collected interferometric data of \obj\ at multiple frequencies with different resolutions, obtained maps of the jet structure and spectral index distribution on mas scales, measured the advancing velocity of the hotspots, and analysed the variability and radio spectrum of the whole source to verify it as a CSO from multiple perspectives.
Throughout this paper, we adopt a flat $\Lambda$CDM cosmological model with H$_{0}=70$\,km s$^{-1}$ Mpc$^{-1}$, $\Omega_{\Lambda}=0.73$, and $\Omega_{\mathrm{m}}=0.27$. In this model, $1$\,mas angular size corresponds to 6.79 pc projected linear size at the source redshift $z=4.56$ \citep{2018ApJS..239...20M}. A proper motion speed of 1 mas yr$^{-1}$ corresponds to $123\,c$. The spectral index $\alpha$ is defined as $S \propto \nu^\alpha$, where $S$ is the flux density and $\nu$ is the observation frequency.

\section{Data and Methods}

VLBI data from five epochs were used for the study in this paper, see Table \ref{tab:obs} for observational details. Four epochs  were obtained from the Astrogeo VLBI database\footnote{Astrogeo database maintained by L. Petrov: \url{http://astrogeo.org/}.}. This database is mainly used for geodesy and astrometry studies \citep[e.g.,][]{2002ApJS..141...13B}. Most of these observations were made in the snapshot mode at 2 and 8 GHz dual frequencies. The phase of the visibility data has been calibrated, so we only need to import the downloaded data into the \textsc{Difmap} software package \citep{1997ASPC..125...77S} for self-calibration, imaging and model fitting. 

Another epoch of data was observed with the Very Long Baseline Array (VLBA) at 8.4 GHz on March 19, 2017 among a project (code: BZ064) of studying the jet structure and proper motion of a high-redshift AGN sample (Zhang YK et al. in prep.). 
The data rate is 2048 Mbits per second: i.e., a total recording bandwidth of 128 MHz; both left- and right-handed polarisation; 2-bit sampling. \obj\ was observed on six scans, which were interleaved with other sources, to improve the \textit{uv} coverage. The cumulative on-source time of \obj\ is 0.5 hour. After the observations were completed, the raw data were correlated in the NRAO correlator at Socorro (New Mexico, USA). Then, we downloaded the correlated visibility data to the compute system of China SKA Regional Centre prototype located at Shanghai Astronomical Observatory \citep{2019NatAs...3.1030A}, for post-calibration and imaging analysis using the VLBI data processing pipeline, which was developed by our group for continuum imaging observations in the standard phase-reference mode. This procedure included manual fringe fitting using a bright fringe finder source (3C~273) to correct for delay and phase errors between different sub-bands; a global fringe fitting to all sources' data to calculate and remove global phase errors; applying the gain solutions obtained from the fringe fitting to the target source; and solving the antenna-based bandpass functions and applying them to the target visibility data. After the amplitude and phase calibrations, the visibility data were averaged in the frequency and time domains, and the \obj\ data were exported to \textsc{Difmap} software package for further self-calibration and imaging. After several iterations of "self-calibration and mapping" cycles, the noise of the image was close to the theoretical value, then the data processing was completed.
The final images are created using natural weights,  obtaining the highest resolutions of 0.8 mas\footnote{1 mas corresponds to a projected size of $\sim$6.8 pc.} at 8.4 GHz and 3.7 mas at 2.3 GHz, respectively. The observation on March 19, 2017 had the longest integration time and the broadest bandwidth, therefore obtaining the lowest image noise of 0.1 mJy beam$^{-1}$.

To quantitatively describe the physical properties of the \obj\ jet, we fit its emission structure on the visibility plane with three Gaussian models (Table \ref{tab:parm}). Based on our experience in processing VLBI survey data, we estimate the uncertainty in flux density to be about 5 per cent. The uncertainty of the astrometry of \obj\ is 0.16 mas, which does not affect the jet kinematics analysis because the jet kinematics is related to the relative position change of the jet components, not the absolute position. 
The position error of the jet component is mainly determined by the image signal-to-noise ratio and the size of the synthesised beam. 
We estimated the errors of jet component size ($\sigma_{\rm d}$) and distance ($\sigma_{\rm R}$) following the formula given in \citet{1999ASPC..180..301F}: $\sigma_{\rm d}= (\sigma_{\rm p} \times d_{\rm comp} )/S_{\rm peak}$ and $\sigma_{\rm R}=0.5 \sigma_{\rm d}$, 
where $\sigma_{\rm d}$ is the uncertainty of the fitted component size $d_{\rm comp}$, $\sigma_{\rm p}$ is the root-mean-square noise measured around its position in the residual map, $S_{\rm peak}$ is peak intensity, and $\sigma_{\rm R}$ is the error of angular separation from the reference point (\textit{i.e.}, the brightest component S in this study, in Fig. \ref{fig:vlbi}).

In addition to the above image data, we collected data points of \obj\ from the NASA/IPAC Extragalactic Database (NED) and from the RATAN-600 radio telescope \citep{2012A&A...544A..25M} to construct the radio spectrum of the entire source, which is shown in Figure \ref{fig:spec}. 
This source has been monitored by the 40-metre radio telescope of the Owens Valley Radio Observatory (OVRO) at 15 GHz \citep{2011ApJS..194...29R}. The OVRO radio lightcurve is used for radio variability analysis. 

\begin{table*}
	\centering
	\caption{Information about the VLBI observations of \obj}
	\label{tab:obs}
	\begin{tabular}{ccccccrccc} 
		\hline
	
		Code & Date       & Frequency & Bandwidth & $B_\mathrm{maj}$ & $B_\mathrm{min}$ & $B_\mathrm{P.A.}$             & Peak              & $\sigma_{\rm rms}$  & Ref.   \\
		     & YYYY-MM-DD & (GHz)     & (MHz)     & (mas)            & (mas)            & \multicolumn{1}{c}{($\degr$)} & (mJy beam$^{-1}$) & (mJy beam$^{-1}$)   &        \\
	 (1)  &(2)  & (3) &(4) &(5) &(6) &(7) &(8) &(9) &(10) \\   
		\hline
        BB023 & 1996-05-15  & 2.3 &  32  & 9.0  & 3.7  & $-$20.8 & 839.0 & 1.0 & 1 \\
        ...   & ... & 8.3 &  32  & 2.1 & 0.9 & $-18.2$ & 304.4 & 0.9 & 1 \\
        BG219e & 2014-08-09 & 2.3  & 128 & 8.2 & 4.3 & 18.8 &737.0 & 1.3&1\\
        ...   & ... & 8.7  & 384  & 2.1 & 1.1 & 20.9 & 337.0 & 0.3 & 1 \\
		BZ064b & 2017-03-19 & 8.4  & 512 & 2.5  & 0.8  & $-$1.6  & 97.9  & 0.1 & 2 \\
        UF001q & 2017-09-18 & 2.3  & 96 & 12.5 & 6.1 & $-$26.8 &763.5 & 1.0&1\\
        ...   & ... & 8.7  & 384  & 3.2 & 1.6 & $-$25.5& 323.6 & 0.9 & 1 \\
        UG002d & 2018-03-26 & 2.3  & 96 & 6.6 & 4.1 & 2.8 &685.9 & 0.9&1\\
        ...   & ... & 8.7  & 384  & 1.8 & 1.1 & 4.2 & 270.1 & 0.36 & 1 \\
		\hline
		\multicolumn{10}{p{14cm}}{\footnotesize{Notes: Col.~1 -- project code; Col.~2 -- observation date; Col.~3 -- observing frequency; Col.~4 -- bandwidth; Col.~5--7 -- major and minor axes of the synthesized beam (full width at half maximum, FWHM) and the position angle of the major axis, measured from north to east; Col.~8 -- peak intensity in the image; Col.~9 --  root-mean-square (rms) noise in the image; Col.~10 -- reference to the corresponding VLBI experiment: 1--Astrogeo database; 2--this paper.}} \\
	\end{tabular}
\end{table*}

\section{Results}
\subsection{Radio morphology} \label{sec:mor}

\obj\ is unresolved in images with arcsec resolutions (Figure \ref{fig:racs}) obtained from the VLA/VLASS (resolution of 2.5 arcsec), VLA/FIRST (5 arcsec), ASKAP/RACS (15 arcsec) and VLA/NVSS (45 arcsec). 
In the 5-GHz VLBA images, the source is clearly resolved into two components along the north-south direction  \citep{2007ApJ...658..203H}, separated by $\sim$8 mas.  These two components show similar morphology, similar to Compact Symmetric Objects  \citep{1998A&A...337...69O,2016MNRAS.459..820T}. 
This 5-GHz VLBA observation was made in 2006 January. Since the source does not have prominent variability (Section \ref{sec:var}), our comparison of the VLBI flux density with the total flux density measured by RATAN-600 at the same frequency in October 2006 reveals that 94 per cent of the radio emission comes from the mas-scale structure, indicating that the large-scale extended emission at GHz frequencies emission does not exist or is too weak to be detected.

\begin{figure}
    \centering
    \includegraphics[width=0.45\textwidth]{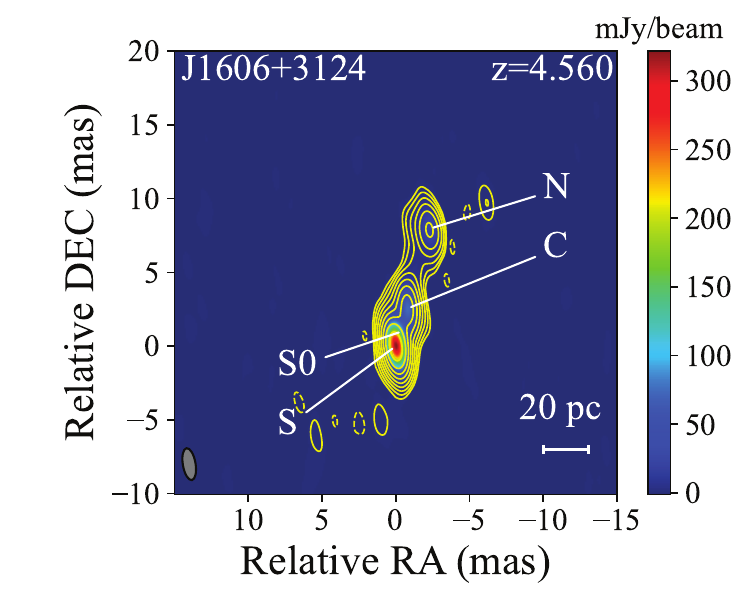}
    \caption{VLBA image of \obj\ at 8.4 GHz. The observation was made on 2017 March 19. The image parameters are referred to Table \ref{tab:obs}. The horizontal bar indicates a 20-pc size scale in projection. The peak intensity is 97.9 mJy~beam$^{-1}$, and the rms noise ($1\sigma$) is 0.1 mJy~beam$^{-1}$. The contours start at three times the rms noise and increase in a step of 2.}
    \label{fig:vlbi}
\end{figure}

Our new VLBA image is shown in Figure \ref{fig:vlbi}. It was observed at 8.4 GHz, with a higher resolution of 2.5 mas $\times$ 0.8 mas than the previous 5-GHz image. Except for two edge-brightened components labelled as N and S previously detected in the 5-GHz VLBA image, another component (C) is found in the intervening region.  The three components are aligned along a position angle of $\sim-17^\circ$, with a total extent of $\sim$56 pc (in projection).

\begin{figure}
    \centering
    \includegraphics[width=0.45\textwidth]{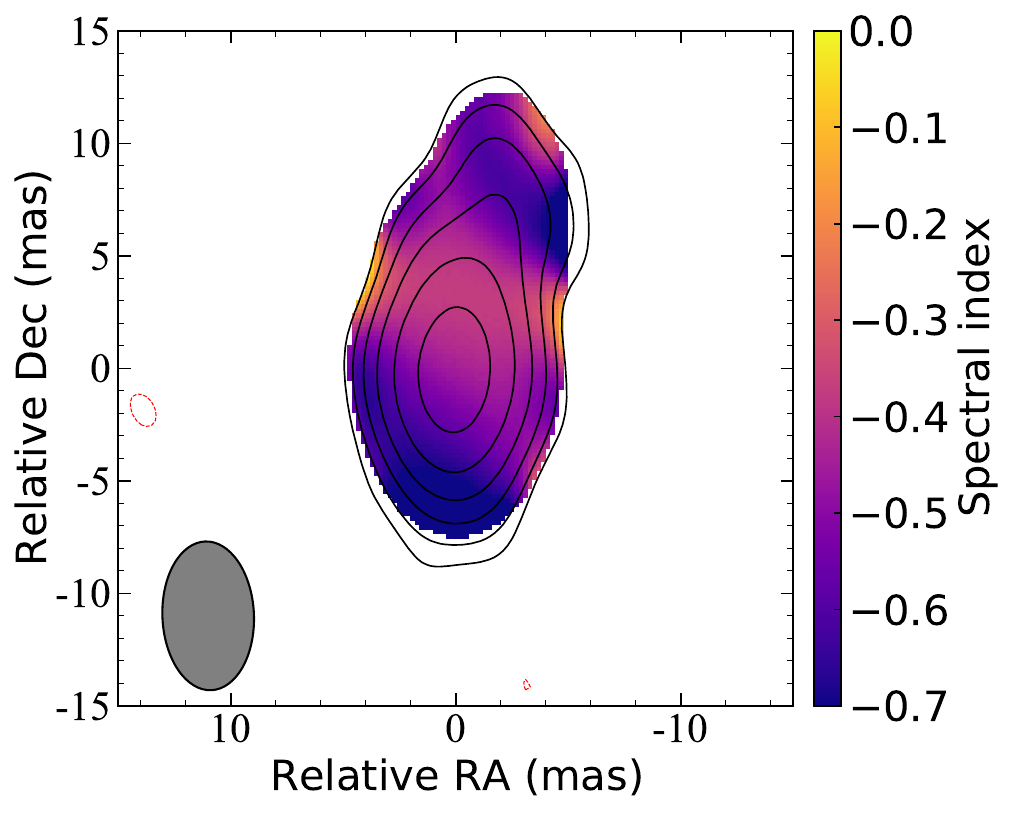}
    \includegraphics[width=0.45\textwidth]{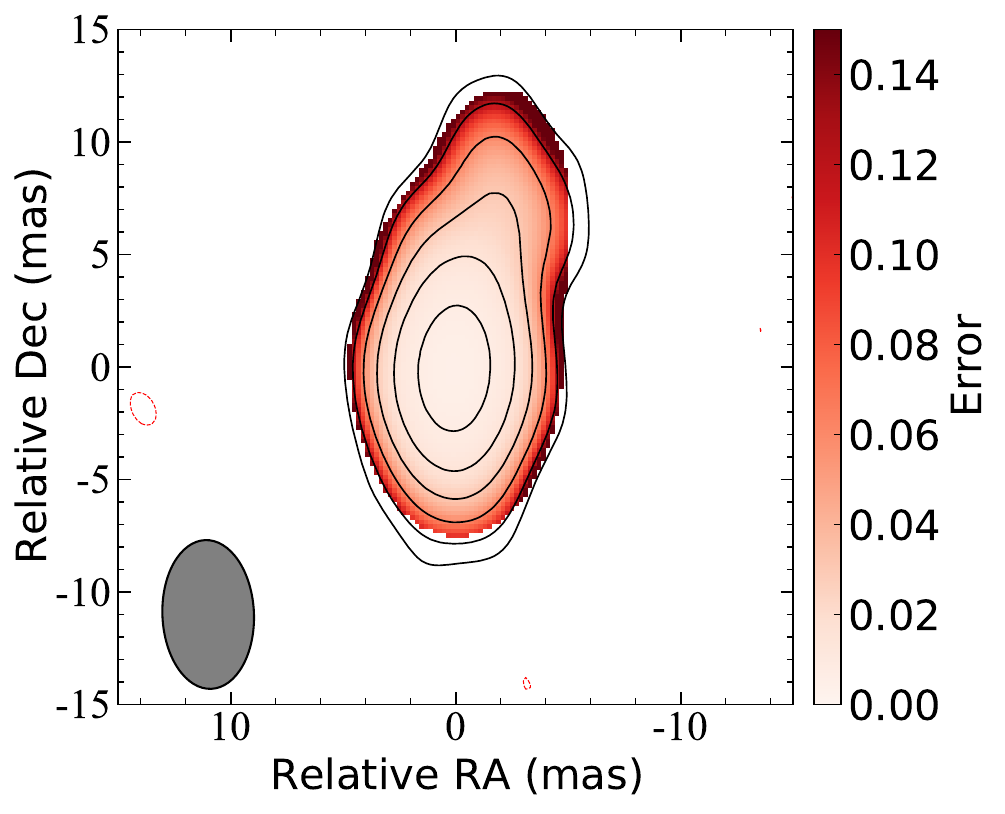}
    \caption{Spectral index map and error map of \obj\ created from the simultaneous 2.3/8.4 GHz VLBI data on epoch 2018 March 26. The contours represent 14 mJy beam$^{-1} \times$ ($-1$, 1, 2, 4, 8, 16, 32).}
    \label{fig:spixmap}
\end{figure}

S is the brightest component at 2, 5 and 8 GHz and is more compact than N. We calculate the spectral indices of S and N for each epoch based on the model fitting results of the simultaneously observed 2 GHz and 8 GHz data, which are $-0.46  \le \alpha_{\rm S} \le -0.66$ and $-1.05 \le \alpha_{\rm N} \le -1.10$.
To investigate the spectral index distribution, we also produced spectral index maps using the 2 and 8 GHz VLBI data. 
We first re-imaged the S-band and X-band data  with uniform image size (1024 pixels) and cell size (0.2 mas/pixel), and used the 'uvtaper' command in \textsc{Difmap} to increase the sensitivity of the extended structure in the X-band image. The two images were then convolved using the same beam (approximately equal to the original beam of the S-band image). Next, the two images were combined using the \textsc{COMB} task in \textsc{AIPS} to generate a spectral index map as well as an error map. 
The alignment of the 2 and 8 GHz images was referenced to the emission peak of the S component. 
Figure \ref{fig:spixmap} shows the spectral index map derived from the 2018 epoch data. Spectral index maps in the other epochs look similar, so we do not repeat the display here.  
Figure \ref{fig:spixmap} clearly shows that the spectrum is flattened toward the central region of the radio structure. 
At both ends of the radio structure, there is a tendency for the spectral index to steepen. The spectral index at the centroid of the southern component (\textit{i.e.}, around the peak position of S) is $\alpha \sim -0.45\pm0.04$, a value between the optically thin and optically thick regimes. The component N obviously has a steep spectrum ($\alpha \sim -0.60\pm0.05$). The spectral indices of N and S inferred from Figure \ref{fig:spixmap} are consistent with those calculated directly from the model fitting results. Although S is the brightest component, its relatively steep spectral index and steepening toward the edge are inconsistent with the flat spectrum of conventionally defined AGN cores. Therefore it is not a position to relate S to the core.

Neither N nor S component satisfies the conventional definition of a flat-spectrum radio core.  Instead, a more natural interpretation is that they are terminal hotspots in the jets.
The radio core may lie in the relatively flatter spectrum region between the N and S components. High-resolution VLBI (\textit{e.g.}, 15 GHz) observations are needed to further confirm the location of the radio core.

\begin{table*}
	\caption{Model-fitting parameters of the VLBI components.}
	\label{tab:parm}
	\begin{tabular}{ccccccccc}
		\hline
		Epoch      &Comp. &  $S_\mathrm{int}$ &  $S_\mathrm{peak}$  & $R$           &  P.A.             &  $d_\mathrm{comp}$       & $\sigma_{\rm d}$        &  $T_\mathrm{B}$\\
		(YYYY-MM-DD)&     &  (mJy) &  (mJy beam$^{-1}$)             &  (mas)        &  ($\degr$)        & (mas)                    & (mas)          & ($\times$10$^{10}$K) \\
		(1)     &(2) & (3)             &(4)              &(5)        &(6)            &(7)                 &(8)         &(9)     \\  
        \hline
        1996-05-15 & S   & 373$\pm$19         & 315$\pm$16          & ...           & ...               & 0.58                     & 0.01           & 10.8$\pm$0.6  \\
                   & C   & 79$\pm$4           & 70$\pm$4            & 2.49$\pm$0.01 & $-160.1\pm$0.3    & 0.39                     & 0.02           & 5.2$\pm$0.5   \\
                   & N   & 49$\pm$2           & 35$\pm$2            & 7.97$\pm$0.02 & $-162.7\pm$0.2    & 0.75                     & 0.03           & 0.84$\pm$0.08 \\
                   
        2014-08-09 & S   & 444$\pm$22         & 383$\pm$19          & ...           & ...               & 0.58                     & 0.01           & 11.7$\pm$0.8  \\
                   & C   & 99$\pm$5           & 78$\pm$4            & 2.60$\pm$0.02 & $-162.3\pm$0.6    & 0.64                     & 0.03           & 2.2$\pm$0.2   \\
                   & N   & 40$\pm$2           & 27$\pm$1            & 8.29$\pm$0.03 & $-163.0\pm$0.4    & 0.87                     & 0.07           & 0.47$\pm$0.08 \\
                   
        2017-03-19 & S   & 347$\pm$17         & 334$\pm$17          & ...           & ...               & 0.44                     & 0.02           & 17.3$\pm$1.4  \\
                   & S0  & 21$\pm$1           & 9$\pm$1             & 1.12$\pm$0.05 & $-155.2\pm$1.4    & ...                        & ...            & ...           \\
                   & C   & 81$\pm$4           & 67$\pm$3            & 2.66$\pm$0.01 & $-162.6\pm$0.6    & 0.53                     & 0.03           & 2.8$\pm$0.3   \\
                   & N   & 34$\pm$2           & 24$\pm$1            & 8.27$\pm$0.02 & $-163.7\pm$0.4    & 0.75                     & 0.06           & 0.58$\pm$0.10 \\
                   
        2017-09-18 & S   & 326$\pm$16         & 328$\pm$16          & ...           & ...               & 0.53                     & 0.01           & 10.4$\pm$0.64  \\
                   & C   & 95$\pm$5           & 63$\pm$3            & 2.49$\pm$0.02 & $-164.3\pm$0.8    & 0.61                     & 0.04           & 2.3$\pm$0.4   \\
                   & N   & 39$\pm$2           & 30$\pm$2            & 8.14$\pm$0.02 & $-163.4\pm$0.3    & 1.16                     & 0.04           & 0.26$\pm$0.02 \\
                   
        2018-03-26 & S   & 297$\pm$15         & 275$\pm$14          & ...           & ...               & 0.47                     & 0.01           & 11.9$\pm$0.7  \\
                   & S0  & 24$\pm$1           & 10$\pm$1            & 1.11$\pm$0.02 & $-177.0\pm$1.3    & ...                        & ...            & ...           \\
                   & C   & 82$\pm$4           & 71$\pm$4            & 2.63$\pm$0.01 & $-163.4\pm$0.6    & 0.57                     & 0.03           & 2.3$\pm$0.3   \\
                   & N   & 36$\pm$2           & 26$\pm$1            & 8.24$\pm$0.02 & $-163.9\pm$0.3    & 0.76                     & 0.04           & 0.56$\pm$0.07 \\ 
        \hline
        
\multicolumn{9}{p{16cm}}{\footnotesize{Notes: Col.~1 -- observing date; Col.~2 -- label of the fitted model component; Col.~3 -- peak intensity; Col.~4 -- total flux density; Col.~5 -- angular separation of the component with respect to S ; Col.~6 -- position angle of the component with respect to S, measured from north to east; Col.~7 -- fitted FWHM size of the circular Gaussian model component. Note: the fit of S0 is degenerated to a point source model. ; Col.~8 -- uncertainty of $D_{\rm comp}$; Col.~9 -- brightness temperature. 
}} \\
	\end{tabular}
\end{table*}

\subsection{Radio spectrum} \label{sec:spec}

\begin{figure*}
    \centering
    \includegraphics[width=0.45\textwidth]{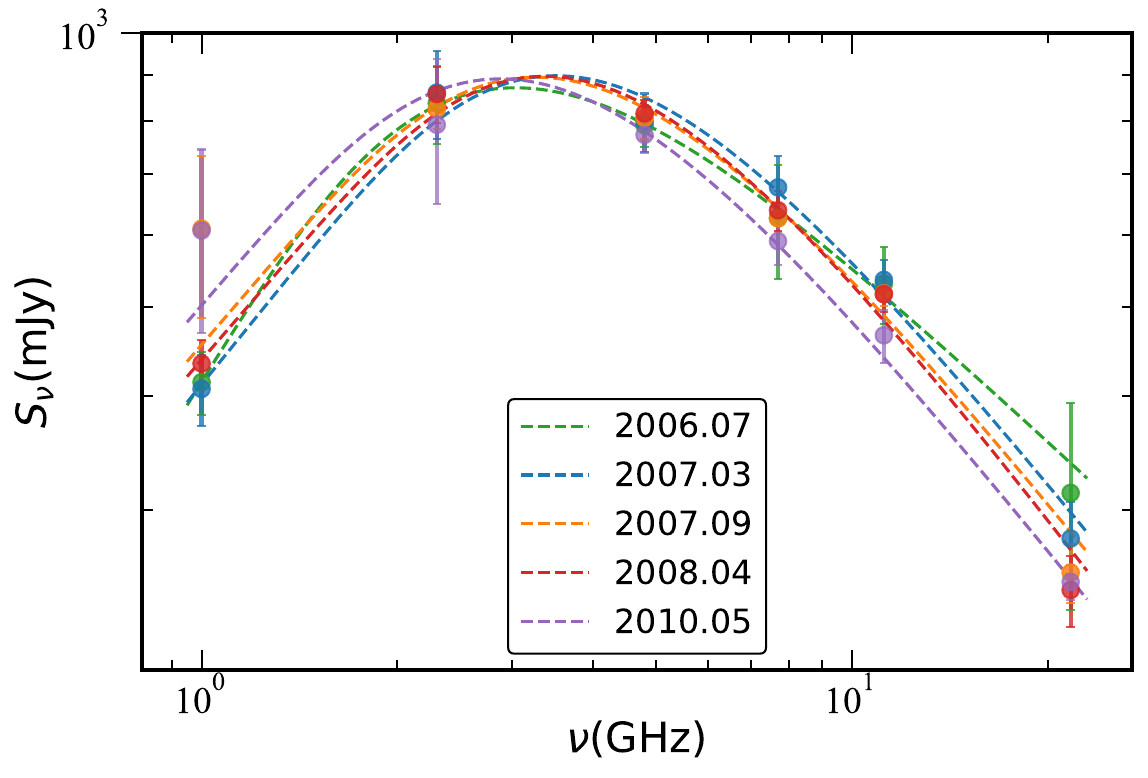}
    \includegraphics[width=0.45\textwidth]{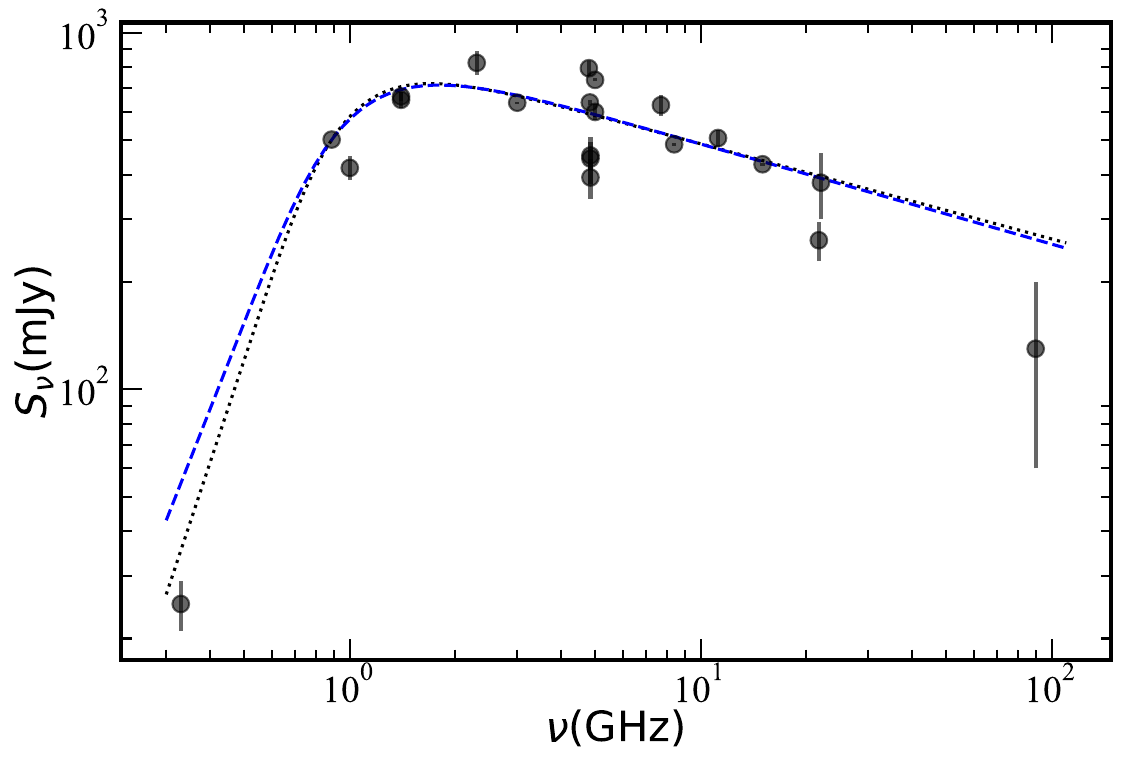}    
\caption{Radio spectrum of \obj. \textit{a}: Spectra obtained from the RATAN-600 data on five epochs.  The dashed lines represent fits with the synchrotron self-absorption model; see Section \ref{fig:spec} for more discussion.  \textit{b}: Spectrum constructed using all data points. The black-colour data points are from the NED database and have been augmented with the averaged flux densities from RATAN data (see Table \ref{tab:sed} for details). The fit represented by the long dashed line adoptes a fixed value of $\alpha_{\rm thick} = 2.5$, while in the fit represented by the short dashed line $\alpha_{\rm thick}$ is a free parameter. }
    \label{fig:spec}
\end{figure*}

\obj\ has been considered as a GHz peaked spectrum source in previous studies \citep{1991ApJ...380...66O,1993ApJS...88....1S,2017MNRAS.467.2039C}. 
Recent RATAN-600 simultaneous multi-band observations have confirmed the peaked radio spectrum of \obj\ \citep{2021MNRAS.508.2798S}. 
Figure~\ref{fig:spec}-a shows the RATAN-600 data in five epochs between October 2006 and May 2010 \citep{2012A&A...544A..25M,2021MNRAS.508.2798S}. The data points in different epochs consistently show a turnover around 3 GHz (in the observer's frame). Since the component S dominates the total flux density at GHz frequencies, the RATAN-600 spectra actually describes the S component's spectra approximately.  The peaked spectrum of such a compact radio source can be attributed to synchrotron self-absorption (SSA). In contrast to \citet{2021MNRAS.508.2798S} who fitted the spectrum with a two-section power-law function, we used the following function to fit the radio spectrum, which describes the self-absorbed synchrotron radiation emitted by electrons with a power-law energy distribution in a homogeneous magnetic field \citep{1970ranp.book.....P,1999A&A...349...45T}:
\begin{equation}
  F_{\nu} = F_{\rm m}(\frac{\nu}{\nu_{\rm m}})^{\alpha_{\rm thick}} \frac{1-\exp(-\tau_{\rm m}(\nu/\nu_{\rm m})^{\alpha_{\rm thin}-\alpha_{\rm thick}})}{1-\exp(-\tau_{\rm m})},
\end{equation}
\begin{equation}
\tau_{\rm m}\approx \frac{3}{2}(\sqrt{1-\frac{8\alpha_{\rm thin}}{3\alpha_{\rm thick}}}-1),
\end{equation}
where $\nu_{\rm m}$ is the turnover frequency; $F_{\rm m}$ is the maximum flux density at the turnover frequency; $\alpha_{\rm thin}$ and $\alpha_{\rm thick}$ describe the spectral indices of the optically thin and thick parts of the spectrum, respectively; $\tau_{\rm m}$ can be approximated as the optical depth at the turnover.
The fitted parameters are listed in Table~\ref{tab:sedfit}. 
It should be noted that the 1-GHz data points of epochs 2007.09 and 2010.05 have large measurement errors, i.e., three times those of the other epochs, leading to less stringent constraints on its spectral shape for the optically thick part. Therefore, we fixed  $\alpha_{\rm thick}=0.83$ (i.e., equivalent to the smallest value in the fitted results of the other epochs) of these two epochs in the fit.
The fitted turnover frequency falls between 2.7 and 3.3 GHz, which is roughly consistent with the result of 2.5 GHz obtained by \citet{2021MNRAS.508.2798S}.
The values of the fitted parameters are all consistent within $2\sigma$, supporting the absence of strong variability and SED variation in \obj, which is consistent with its CSO identification.

In Figure~\ref{fig:spec}-b, we collected the data points from the NED database based on their availability (see Table \ref{tab:sed}). Two data points obtained from the latest VLA/VLASS and ASKAP/RACS observations are also added to the plot. The mean values at every RATAN-600 frequencies are also added (Table~\ref{tab:sed}). Although these observations were made at different epochs and resolutions, the overall radio spectrum still provides useful information given that the flux density of this source is concentrated on the parsec scale and has no rapid and significant variability (Section \ref{sec:var}). 
Compared to the RATAN spectra, the shape of the low-frequency section of the NED spectrum is more strongly constrained due to the fact that there are more data points below 2 GHz (i.e., at 0.33, 0.888, 1, and 1.4 GHz). This results in the NED spectrum has a higher optically thick spectral index $\alpha_{\rm thick}$ and a lower $\nu_{\rm m}$ than the fits in Figure~\ref{fig:spec}-a. 

The derived optically-thick spectral index $\alpha_{\rm thick} = 3.0$ in the NED spectrum is slightly larger than the classical SSA spectral index $\alpha_{\rm SSA} = 2.5$, which may be due to a less accurate fit to the optically-thick spectrum shape resulting from the lack of simultaneously observed data points in the low-frequency band, or to the fact that the low-frequency absorption results from a mixture of SSA and free-free absorption (FFA).
Future simultaneous multi-band (covering low frequencies below 1 GHz)  radio data are crucial to tightly constrain the turnover frequency, low-frequency spectral index $\alpha_{\rm thick}$ and the associated astrophysical mechanisms. 
The fitted turnover frequency ($\nu_{\rm m}$) corresponds to 10.6--17.8 GHz in the source's rest frame, classifying \obj\ as a high-frequency peaker \citep[HFP,][]{2000A&A...363..887D}, a heterogeneous class of CSOs and blazars \citep{2003PASA...20...79D}. The observed small size and symmetric radio structure make it more inclined to be a CSO. The source size and $\nu_{\rm m}$ of \obj\ are consistent with the anti-correlation between the source-frame peak frequency and linear size of CSO and medium-sized symmetric objects \citep{1990A&A...231..333F,1998PASP..110..493O,2009AN....330..120F}, suggesting that the smaller sources are younger and have higher turnover frequencies.

Another interesting feature in Figure \ref{fig:spec}-b is a possible drop between 22 and 90 GHz (122 and 500 GHz in the source rest frame, respectively). 
This phenomenon could be associated with the ageing of synchrotron-emitting electrons due to the radiative loss. The main sources of radiative loss are synchrotron radiation and inverse Compton scattering.
Since the energy density of the cosmic microwave background (CMB) increases with $(1+z)^4$, the contribution of cooling by relativistic electrons scattering CMB photons becomes important in the high-redshift Universe and may be responsible for the observational absence of large-scale extended jets \citep{2014MNRAS.438.2694G}. 
However, in a compact pc-scale radio source (as is the case in this paper), the equipartition magnetic field strength far exceeds that of the CMB (Table~\ref{tab:Bfield}), leading to that the synchrotron loss is dominant in compact jets.

\begin{table}
    \centering
    \caption{Radio flux densities of \obj\ used for spectrum fitting.} 
    \begin{tabular}{cccc} \hline\hline
    $\nu$ (GHz) & telescope & $S$ (mJy)  & ref. \\ \hline 
   90.0    & NRAO11m  & 130$\pm$70 & 1 \\
   22.0    & KVN      & 380$\pm$80 & 2 \\
   21.7    & RATAN-600& 262$\pm$33 & 3 \\
   15.0    & OVRO     & 428        & 4 \\
   11.2    & RATAN-600& 507$\pm$28 & 3 \\
    8.4    & VLA      & 486.8      & 5 \\
    7.7    & RATAN-600& 627$\pm$43 & 3 \\
    5.0    & VLA      & 600$\pm$30 & 6 \\
    5.0    & VLBA     & 738.1      & 7 \\
    4.85   & GBT91m   & 444$\pm$67 & 8 \\
    4.85   & GBT91m   & 453$\pm$40 & 9 \\
    4.85   & GBT91m   & 393$\pm$51 & 10 \\
    4.83   & GBT91m   & 638        & 11 \\
    4.8    & RATAN-600& 796$\pm$39 & 3 \\
    3.0    & VLA/VLASS& 637$\pm$4  & 12 \\
    2.3    & RATAN-600& 824$\pm$63 & 3 \\
    1.4    & VLA/NVSS & 663$\pm$20 & 13 \\
    1.4    & VLA/FIRST& 649$\pm$32 & 14\\
    1.0    & RATAN-600& 418$\pm$32 & 3 \\ 
    0.888  &ASKAP/RACS& 502$\pm$1  & 15 \\
    0.33   &Westerbork&  25$\pm$4  & 16 \\
    \hline 
    \end{tabular} \\ 
    References: 1 -- \citet{1977AJ.....82..776O}; 2 -- \citet{2017ApJS..228...22L}; 3 -- \citet{2012A&A...544A..25M} (here, average values of all epochs are presented.); 4 -- \citet{2014MNRAS.438.3058R}; 5 -- \citet{2007ApJS..171...61H}; 6 -- \citet{1978AJ.....83..685O}; 7 -- \citet{2007ApJ...658..203H}; 8 -- \citet{1991ApJS...75....1B}; 9 -- \citet{1996ApJS..103..427G}; 10 -- \citet{1991ApJS...75.1011G}; 11 -- \citet{1990ApJS...72..621L}; 12 -- \citet{2020RNAAS...4..175G}; 13 -- \citet{1998AJ....115.1693C}; 14 -- \citet{1995ApJ...450..559B}; 15 -- \citet{2020PASA...37...48M}; 16 -- \citet{1997A&AS..124..259R}. 
    \label{tab:sed}
\end{table}

\begin{table}
    \centering
    \caption{ The spectrum fitting parameters.} 
    \begin{tabular}{ccccc} \hline\hline
    Epoch     & $F_{\rm m}$            & $\nu_{\rm m}$    & $\alpha_{\rm thick}$ & $\alpha_{\rm thin}$      \\
    YYYY-MM   & (mJy)         & (GHz)        &                      &               \\
    (1)       & (2)           & (3)          & (4)                  & (5)            \\
    \hline 
    \multicolumn{5}{c}{RATAN-600 spectrum} \\
    2006.07   & 870$\pm$19    & 3.0$\pm$0.1  & 1.2$\pm$0.2          &$-0.68\pm0.07$   \\
    2007.03   & 896$\pm$68    & 3.3$\pm$0.4  & 0.88$\pm$0.27        &$-0.91\pm0.15$   \\
    2007.09   & 893$\pm$76    & 3.1$\pm$0.4  & 0.83                 &$-0.91\pm0.21$    \\
    2008.04   & 894$\pm$58    & 3.2$\pm$0.3  & 0.83$\pm$0.25        &$-0.96\pm0.15$    \\
    2010.05   & 889$\pm$79    & 2.7$\pm$0.3  & 0.83                 &$-0.91\pm0.11$    \\ \hline
    \multicolumn{5}{c}{NED spectrum} \\
    ...       & 721$\pm$34    & 1.7$\pm$0.1  & 3.0$\pm$0.7          &$-2.7\pm0.1$      \\
    ...       & 714$\pm$29    & 1.8$\pm$0.1  & 2.5                  &$-2.8\pm0.1$       \\
        \hline 
    \end{tabular} \\ 
    Notes: Col. 1 -- observing date; Col. 2 -- maximum flux density at the turnover frequency; Col. 3 --  turnover frequency; Col. 4 -- spectral index of the optically thick part in the spectrum; Col. 5 -- spectral index of the optically thin part in the spectrum. Note: the optically thick part of the RATAN spectrum is not well constrained in epochs 2007.09 and 2010.05.    
    \label{tab:sedfit}
\end{table}

\subsection{Variability} \label{sec:var}

\begin{figure*}
    \centering
    \includegraphics[width=0.9\textwidth]{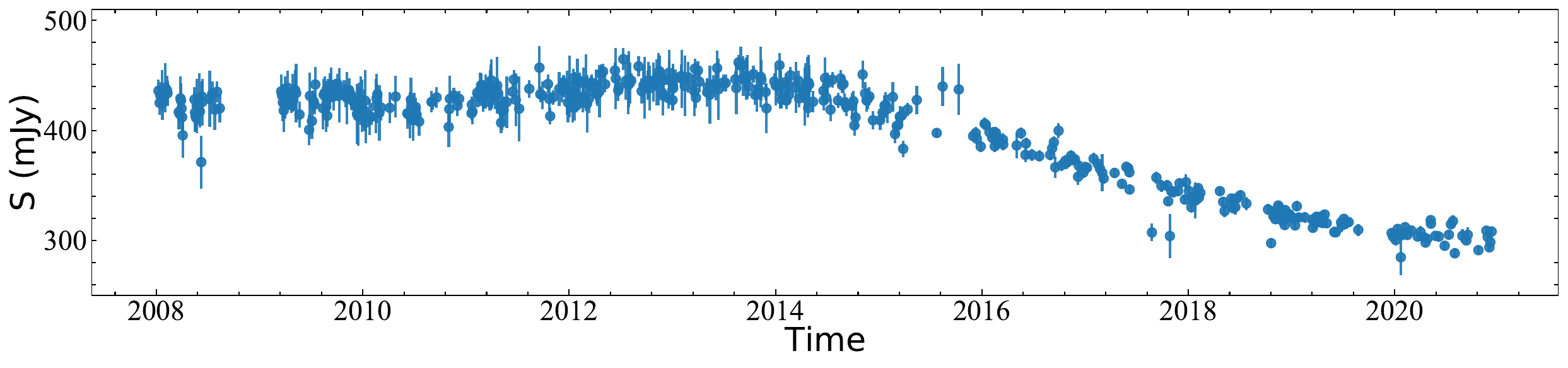}
    \caption{Lightcurve of \obj\ observed by the 40-metre radio telescope of the Owens Valley Radio Observatory at 15 GHz. 
    }
    \label{fig:var}
\end{figure*}

Figure \ref{fig:var} shows the 15-GHz lightcurve of \obj. Except for two gaps from epoch 2008.6 to epoch 2009.2 and from 2019.6 to 2020, the monitoring program continuously covers the period from 2008.0 to 2020.9. The average time interval between two adjacent observations is approximately six days. Some bad data points that deviate significantly from the mean value or show large measurement errors were discarded. 

We used the modulation index $V$ to characterise the variability in the lightcurve \citep{2021arXiv210806039M}. 
\begin{equation}\label{e_var2}
V = \frac{1}{\overline{S}}\sqrt{\frac{N}{N-1} (\overline{S^2} - \overline{S}^2)},
\end{equation}
\begin{equation}\label{e_var1}
\eta = \frac{N}{N-1}\left( \overline{w S^2} - \frac{\overline{w S}^2}{\overline{w}}\right),
\end{equation}
where $N$ is the number of data points, $S$ is the flux density, $\overline{S}$ is the mean of $S$, $w_i$ is the weight of the $i$th data point denoted by $w_i=1/\sigma_i^2$, $\sigma_i$ is the measurement uncertainty, and $\eta$ is a measure of the statistical significance of the variability.  The statistical significance of  variability is described by the parameter $\eta$. In \citet{2021arXiv210806039M}, highly variable sources are defined as those with parameters $V >0.51$ and $\eta >5.53$. 

There is no significant variability with the modulation index $V=0.03$ and $\eta=1.5$ over a time scale of 5.8 years from 2008.6 to 2014.4, only showing a slow rise since 2011.6 and reaching a flat plateau between 2012.3 and 2014.0. After 2014.4, the flux density gradually decreased until it reached a minimum of $\sim$300 mJy beyond 2020. The modulation index $V$ for the period 2014.4--2020 is 0.11, which implies a weak variability, and the statistical confidence $\eta$ for this weak variability is 42. The VLBI data points show a decrease in the flux density from 2014 to 2018 at 8.4 GHz frequency, consistent with the slow decline revealed by the 15-GHz lightcurve. After 2020, \obj\ became a non-variable source according to the variability metrics $V=0.03$ and $\eta=2.8$. Therefore, in general, \obj\ can be classified as a weakly variable or non-variable source.  We also checked other variability metrics \citep[such as,][]{2008A&A...485...51H}, which also point to the non-variability of \obj. The variability property of \obj\ is consistent with its CSO galaxy classification \citep{2001AJ....122.1661F,2010MNRAS.402...87A}, however, in contrast to the rapidly variable beamed AGN (i.e., blazars) \citep{2017MNRAS.468...69Z,2020NatCo..11..143A}.

\subsection{Proper motion} \label{sec:pm}

\begin{figure}
    \centering
    \includegraphics[width=0.45\textwidth]{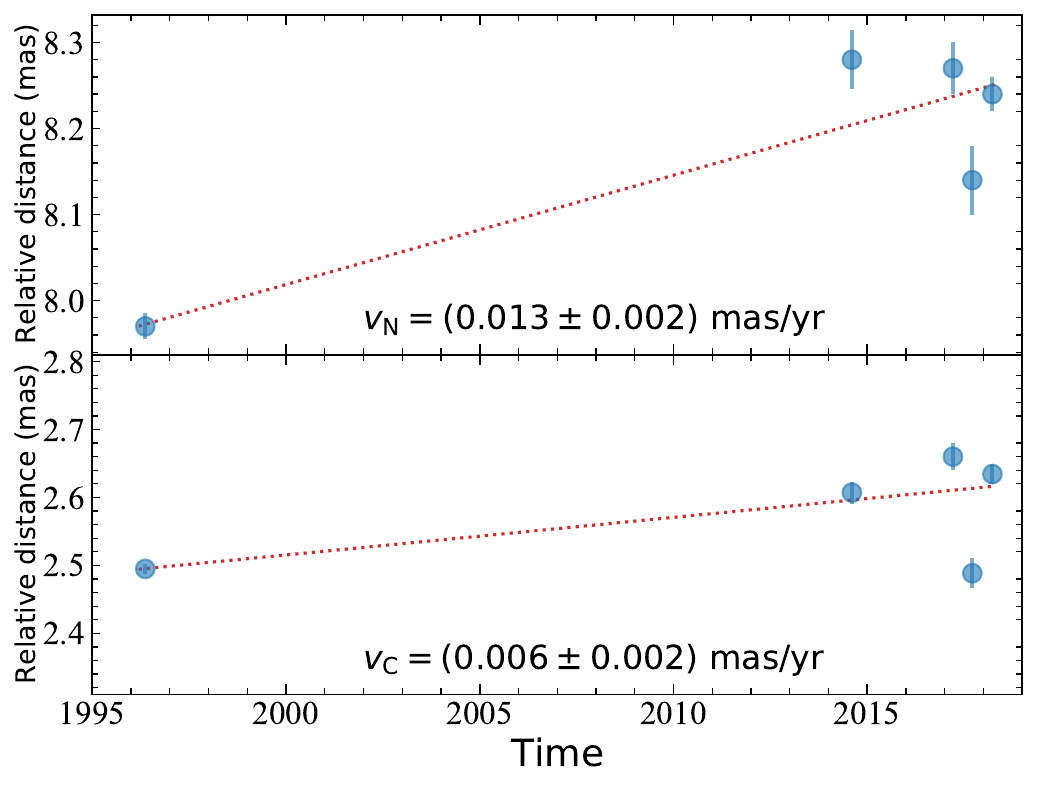}
    \caption{Expansion rates of VLBI components. $\mu_{\rm N-S}:  0.013\pm0.002$ mas yr$^{-1}$; $\mu_{\rm C-S}: 0.006\pm0.002$ mas yr$^{-1}$.}
    \label{fig:pm}
\end{figure}

The current VLBI data do not yet allow the location of the radio core to be determined, therefore we turn to calculate the separation velocity of the VLBI component with respect to the brightest terminal component S. 
Using a linear regression fit shown in Figure~\ref{fig:pm}, we obtained the expansion velocities of N and C with respect to S, $\mu_{\rm N-S}$: $0.013\pm0.002$ mas yr$^{-1}$ and $\mu_{\rm C-S}$: $0.006\pm0.002$ mas yr$^{-1}$, corresponding to apparent transverse speeds of $1.60 \pm 0.25\, c$ (N) and $0.74 \pm 0.25\, c$ (C).  S0 was detected only in two epochs separated by $\sim$3.5 years ($\sim$ 1.0 year in the source rest frame), which is too short to produce a reliable proper motion result.
The derived jet proper motions of \obj\ are significantly lower than those of the high-redshift blazars in previous studies \citep{2010A&A...521A...6V,2015MNRAS.446.2921F,2018MNRAS.477.1065P,2020SciBu..65..525Z,2020NatCo..11..143A}, but consistent with the hotspot separation speed in CSOs \citep{ 2005ApJ...622..136G,2012ApJS..198....5A,2012ApJ...760...77A}. 

Both N and C have positive separation speeds with respect to S, indicating that they are moving away from S. The core, albeit not clearly detected in the VLBI images, should be located between components C and S.
Adopting the simplifying assumption that the hotspots expand at the same rate at both sides and a constant rate during its early growth, the derived apparent advancing velocity of the hotspot is $\beta_{\rm app}\sim0.8$. It constrains the jet intrinsic velocity ($\beta$, in unit of the speed of light) to between 0.6 and 0.8 (Figure~\ref{fig:app_v}). The core possibly lies between C and S, then the arm length ratio of the advancing jet to the receding jet places a lower limit of the jet viewing angle of $\gtrsim 28^\circ$. The suggesting a mildly relativistic jet flow at a moderate viewing angle. 
\section{Discussion}

\subsection{Radio source classification}

From the radio images alone, \obj\ could be a one-sided core-jet source or a compact symmetric object. The brightness temperature can be calculated from the VLBI observables (Table~\ref{tab:parm}), e.g., component size and flux density: $T_{\rm B} = 1.22 \times 10^{12} (1+z) S/(d_{\rm comp}\nu)^2  \,K$, where $S$ is the flux density of the fitted Gaussian component in Jy, $d_{\rm comp}$ is the full width half maximum of the Gaussian component in mas, and $\nu$ is the observing frequency in GHz.  Evidence in support of a core-jet source comes from the identification of component S as the radio core, which has the highest brightness temperature exceeding the  inverse Compton limit \citep{1969ApJ...155L..71K}. The one-sidedness of the jet is usually due to the Doppler boosting effect; however, other observational evidence seem not to support a highly relativistic jet. The spectral index distribution (steep spectra at both edges and a flat spectrum in the middle), the overall peaked spectrum, the slow long-term flux density variation, and the mildly relativistic jet speed all point to the fact that \obj\ is a CSO. In the framework of CSO model, N and S are two terminal hotspots; the central core is not prominent, which is common in CSO galaxies \citep{2000ApJ...534...90P}. 
The only two points that are not favourable for CSO identification are the high brightness temperature (Table~\ref{tab:parm}) of the terminal hotspot S (slightly higher than the canonical value of the equipartition brightness temperature limit, \citealt{1994ApJ...426...51R}) and the brightness asymmetry of S and N (brightness ratio $\gtrsim$10, which is larger than those of conventional CSO sources, \citealt{2000ApJ...534...90P}). Such high brightness temperatures could be related to the extreme compactness of the hotspot and the high magnetic field strength.

This asymmetric brightness of N and S obviously cannot be explained by the Doppler beaming effect, which would make the advancing jet brighter; however this contradicts the VLBI images: the northern (advancing) jet is longer but weaker than that of the southern (receding) jet. 
A likely explanation is that the two-sided jets encounter different interstellar medium (ISM) environments, with the southern jet encountering a more dense medium and more intense jet-ISM interactions. This reflects the complexity of the AGN host galaxy environment. The study of more high-redshift AGN will help determine whether this asymmetry or inhomogeneity is more pronounced and more common in the high-redshift AGN.

\subsection{A young radio source}

There are two widely discussed models for the interpretation of the compact radio structure of CSOs. The 'youth' model \citep[e.g.,][]{1982A&A...106...21P} considers the CSOs to be only $10^{4-5}$ years old and at a very early stage of the radio source evolution \citep[e.g.,][]{2012ApJ...760...77A}. 
The 'frustration' model suggests that the host galactic environment of CSOs severely limits the development of its jets, causing them to 
be confined to sub-kpc scale \citep[e.g.,][]{1984AJ.....89....5V,1991ApJ...380...66O}. Measuring the age of a CSO source is a straightforward way to resolve the above controversy. Assuming that the hotspots expand at the same rate at both sides and a constant rate, we can estimate the kinematic age by measuring the ratio of the angular distance between two terminal hotspots to the separation velocity. 
That infers a kinematic age of $\sim$3600 yr (in the source rest frame) and we further estimate the jet ejection time to be around epoch 1420, which supports to the notion that \obj\ is a young radio source.
Because of the possibility that the southern jet is obstructed by the ISM (see discussion in Section 4.1), the true jet length may be longer than the observed one, resulting in that the estimated kinematic age can be considered a lower limit, but should not be two times longer.

Another way to estimate the age of the synchrotron source is to measure the lifetime of the synchrotron electrons (also known as the spectral age) from the truncation of the radio spectrum at millimetre wavelengths. The spectral age depends on the magnetic field strength and the spectral break frequency \citep{2009MNRAS.395..812M}.
The NED spectrum in Figure \ref{fig:spec} implies a possible truncation between 22 and 90 GHz (122--500 GHz in the source rest frame). Future high-frequency ($>22$ GHz) observations are needed to precisely constrain the spectral break frequency and thus obtain the synchrotron age in combination with the magnetic field strength estimated from the energy equipartition assumption.

\subsection{High-redshift dust-obscured AGN}

High-redshift AGN, especially those near the end of the cosmic reionisation, have received significant attention because they provide strong constraints on the growth of the earliest SMBHs \citep{2010A&ARv..18..279V,2021Galax...9...23S}.
However, the search for high-$z$ AGN poses some technical challenges, mainly the difficulty of obtaining the spectroscopic redshifts of high-$z$ objects.  The presence of large amounts of gas and dust in the host galaxies of high-$z$ AGN obscures the emission from the central AGN, further increasing the difficulty of detection. 
By cross-matching the Sloan Digital Sky Survey (SDSS) and NASA's Wide-field Infrared Survey Explorer (WISE) catalogues, \citet[][]{hickox2017} found that a WISE colour of $W2-W3 = 3.3$ mag clearly separates the obscured AGN from the unobscured ones. The WISE colours $W2-W3$ and $W1-W2$ of \obj\ are 3.81 mag and 0.37 mag, respectively \citep[][]{2013wise.rept....1C}, clearly classifying it as a dust-obscured source. Most dust-obscured high-redshift infrared AGN are found to be radio weak, such as WISE $J224607.56-052634.9$ \citep[$z = 4.6$,][]{2020ApJ...905L..32F} and COS-87259 \citep[$z=6.8$,][]{2021arXiv210801084E}, whose radio emission could be a mixture of weak jet, quasar-driven winds, and star formation activity \citep{2021MNRAS.506.3641G,2021arXiv210607783R}. High-redshift galaxies with extremely high power jets like \obj\ are rare.

\citet[][]{bicknell1997} predicted that the host galaxies of compact and young AGNs harbour gas-rich medium, with a total cold gas mass up to $\rm 10^{10} - 10^{11} \ M_\odot$.
J1606+3124 is classified as a CSO source, and its WISE colour implies dust-obscuration,  which indicate that this young AGN is embedded in a dust- and gas-rich ambient medium.  
In compact radio sources, the HI absorber has a high gas covering factor, resulting in a high detection rate for HI absorption \citep[e.g.,][]{curran2013,aditya2018}. Future HI absorption observations of \obj, in combination with IR observations, can be used to estimate the percentage of atomic and molecular gas in the neutral gas.

\subsection{A CSO in high-redshift Universe}

The projected size of \obj\ is only 56 pc 
and it is located at the beginning of the high-power-jet sequence in the Power-Size (P-D) diagram describing the dynamical evolution of the extragalactic radio sources \citep{1997AJ....113..148O,2010MNRAS.408.2261K,2012ApJ...760...77A}. In the 'radio power-hotspot velocity' diagram shown in Figure 4 of \citet{2012ApJ...760...77A}, \obj\ belongs to a group with the high radio power and high hotspot advancing velocity, which indicates that the total jet power of \obj\ is intrinsically very high, making it stand out in the cosmic era at $z=4.56$. If the central AGN can remain active for a sufficiently long time, the source has a good chance to grow into a Fanaroff-Riley type II (FRII) galaxy \citep{1974MNRAS.167P..31F}. It is worth noting that the number of extended FRII galaxies detected in radio surveys at redshifts above 4 is very small. One possible reason  is that the relativistic electrons in the high-$z$ extended radio jets or lobes are likely to lose a significant amount of energy due to the inverse Compton scattering of CMB photons, resulting in that they are too weak to be detected by current radio telescopes \citep{2014MNRAS.438.2694G}. These factors would make that a significant fraction of HzRGs are young CSO sources \citep{2004NewAR..48.1157F,2008MNRAS.388.1335W,2017MNRAS.469.4083S}. 
Low frequency (e.g., below 300 MHz) radio interferometers have the potential to recover these "missing" extended radio galaxies at high redshifts \citep{2014MNRAS.438.2694G,2015aska.confE..71A}.

In addition to \obj, another high-redshift CSO with VLBI observations is FIRST J1427385+331241 \citep[$z=6.12$,][]{2008A&A...484L..39F}. Both sources have compact radio structure, typical CSO-type radio spectra and young age, while the difference between them is that FIRST J1427385+33124 is a quasar, while \obj\ is identified as a galaxy at a relatively lower redshift. For each detected source with jet beaming towards the observer, there must exist other $2\Gamma^2$ ($\Gamma$ is Lorentz factor) misaligned AGN \citep{2014MNRAS.438.2694G}. However, the observed number of misaligned AGN at high redshifts is much lower than predicted. This contradiction remains an open question.

Searching for extended emission of \obj\ can help clarify whether the radio structure we see in VLBI images is a primordial young source or originates from reactivation of the AGN, while the latter mechanism can constrain the duty cycle of the high-$z$ AGN \citep{2015aska.confE.173K}.  High-resolution low-frequency VLBI, such as the International LOFAR Telescope (ILT) \citep{2021arXiv210807290S} and the Square Kilometre Array low frequency (SKA-low) under construction, has the potential to detect the extended emission \citep{2015aska.confE.173K}, if it is indeed present.

\section{Summary and Conclusions}

We have analysed high-resolution VLBI images, multi-band radio spectra and 15-GHz radio variability of \obj.
It has been tentatively identified as a radio galaxy at a redshift of 4.56 (a tenth of the current age of the Universe),
and is an excellent template for studying the triggering mechanism of the earliest SMBH's activity.
All radio observations, including compact triple morphology, peaked radio spectrum at GHz frequencies and slow variability, consistently indicate that \obj\ is a compact symmetric object. The jet kinematics indicates a mildly relativistic jet at a moderate viewing angle, which supports the identification as a radio galaxy. 
Future higher-resolution VLBI imaging observations will help to determine the location of radio core and to constrain the jet velocity and geometry accurately.

Infrared observations suggest a large amount of dust in the galactic nuclear region of \obj. Radio wave is free of dust obscuration, and high-resolution interferometric observations are able to image the nuclear regions of distant AGN with sub-arcsec to mas resolutions and study the feedback of the AGN jets on host galaxies in the early Universe. By measuring the separation velocity and distance between two terminal hotspots, we estimate the age of \obj\ to be $\gtrsim$3600 years, consistent with the classification of a young radio source. With a radio luminosity of $\sim 1.5 \times 10^{29}$~W~Hz$^{-1}$ at 1.4 GHz, \obj\ is among the highest luminosity family of all radio sources \citep{1999ApJ...518L..61V,2018MNRAS.480.2733S}. \obj\ has a high-power and moderately relativistic jet, even in the presence of large amounts of gas in the host galaxy, it still has the chance to evolve into a large-scale FRII galaxy if the AGN activity can be sustained for a considerable time period. 
Observations below 1 GHz using the uGMRT and LOw Frequency ARray (LOFAR) can help to better constrain the low-frequency spectrum shape -- whether it can be described with synchrotron self-absorption alone. On the other hand, adding data points in the high-frequency band of 22--90 GHz is essential for estimating the lifetime of synchrotron relativistic electrons (i.e., the spectral age).

Current radio observations of distant objects can only detect unusually bright AGN (such as \obj) and massive BHs that accrete near the Eddington limit \citep[reviewed by][]{2008A&ARv..15...67M}. The Square Kilometre Array (SKA) and Next Generation Very Large Array (ngVLA) are expected to detect a large number of weak radio sources at high redshifts, crucial for improving our understanding of the co-evolution of SMBHs and galaxies in the early Universe.

\section*{Acknowledgements}

The VLBI data processing made use of the compute resource of China SKA Regional Centre prototype under the financial support by the National Key R\&D Programme of China (2018YFA0404602,2018YFA0404603) and the Chinese Academy of Sciences (CAS, 114231KYSB20170003).
TA thanks the grant support by the Youth Innovation Promotion Association of CAS.
The VLBA experiment BZ064 is sponsored by Shanghai Astronomical Observatory through the MoU with the NRAO.
The authors acknowledge the use of Astrogeo Center database maintained by L. Petrov.
This research has made use of data from the OVRO 40-m monitoring program \citep{2011ApJS..194...29R} which is supported in part by NASA grants NNX08AW31G, NNX11A043G, and NNX14AQ89G and NSF grants AST-0808050 and AST-1109911.
This work has made use of the NASA Astrophysics Data System Abstract Service, and the NASA/IPAC Extragalactic Database (NED), which is operated by the Jet Propulsion Laboratory, California Institute of Technology, under contract with the National Aeronautics and Space Administration.
The National Radio Astronomy Observatory are facilities of the National Science Foundation operated under cooperative agreement by Associated Universities, Inc. 

\section*{Data Availability}
Some VLBI data in this work has been obtained by the Astrogeo database, \url{http://astrogeo.org}. The NASA/IPAC Extragalactic Database (NED: \url{http://ned.ipac.caltech.edu/}) offers the SED data. The remaining data sets are available from the corresponding author with reasonable request.


%

\bibliographystyle{mnras}
\bibliography{GPS} 




\clearpage 
\appendix
\section{Arcsec-scale images}

\begin{figure}
    \centering
    \includegraphics[width=0.45\textwidth]{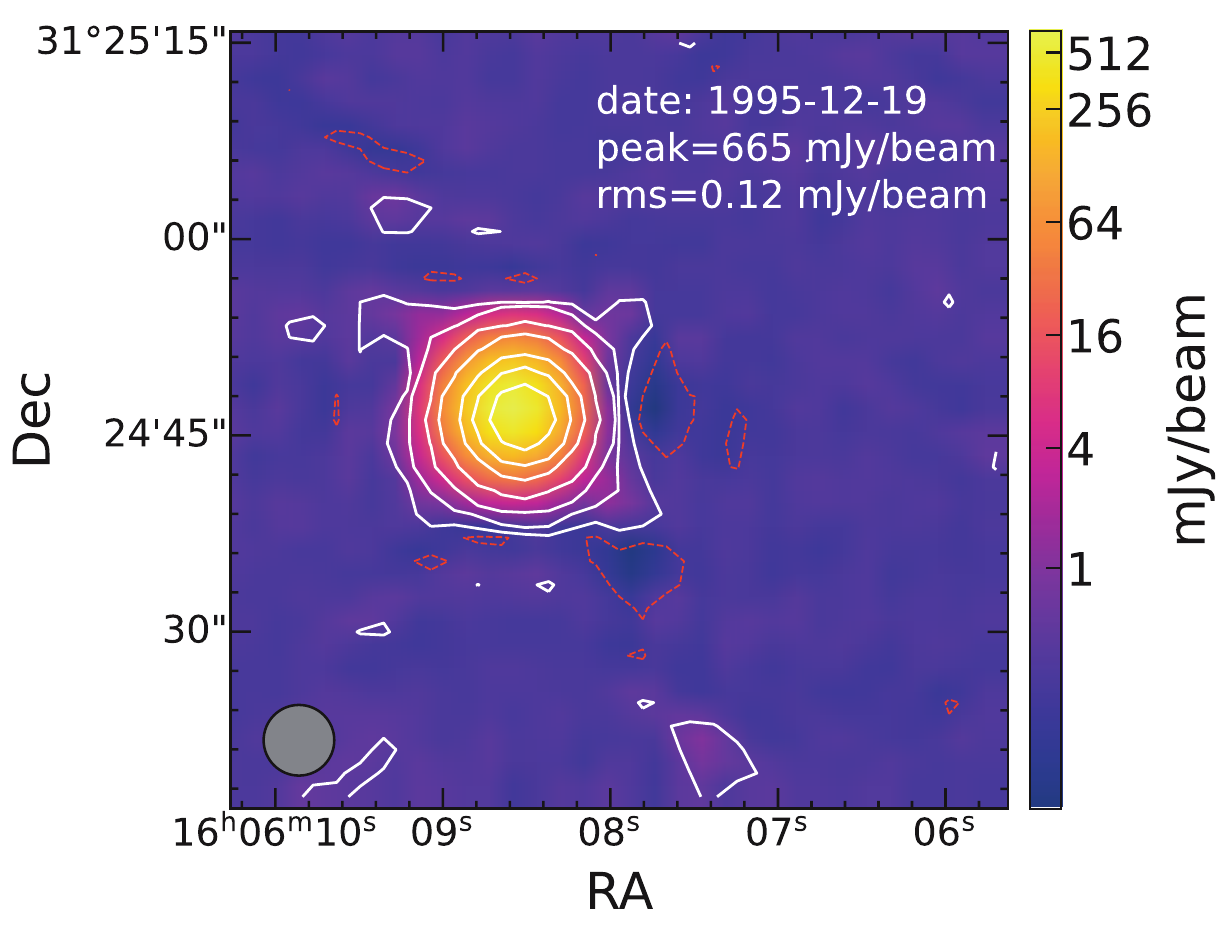} \includegraphics[width=0.45\textwidth]{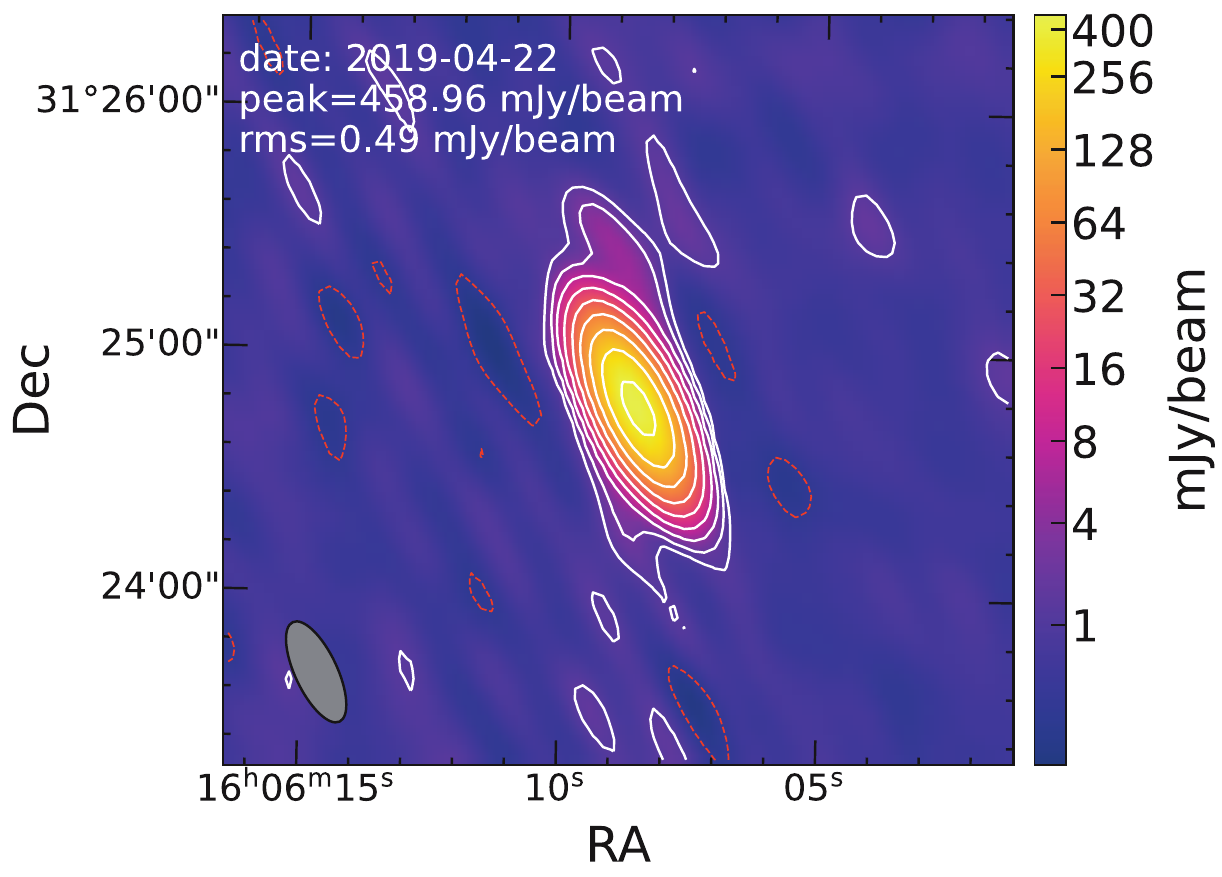} 
    \caption{Arcsec-scale images of \obj. \textit{Top}: FIRST image at 1.4 GHz. The resolution is 5\arcsec. \textit{Bottom}: ASKAP RACS image at 888 MHz. The resolution is 27.1\arcsec\ $\times$ 10.4\arcsec. The lowest contour in the image represents 3 times the rms noise, and the contours increase in steps of 2. The color scale represents the brightness in logarithmic scale.}
    \label{fig:racs}
\end{figure}

\obj\ remains unresolved in arcsec-resolution images. Figure \ref{fig:racs} shows two representative arcsec-scale images obtained from the Faint Images of the Radio Sky at Twenty Centimeters (FIRST) of the Very Large Array at 1.4 GHz \citep{1995ApJ...450..559B} and the pilot continuum survey of the Australian Square Kilometre Array Pathfinder (ASKAP) at 0.888 GHz \citep[the first large-area survey named RACS, ][]{2020PASA...37...48M}.

\section{The magnetic field}

Assuming that the energy density of the emitting particles in the radiation source is close to equipartition with the energy density of the magnetic field, one can estimate the magnetic field strength (usually referred to as the equipartition magnetic field strength) \citep{1970ranp.book.....P}:
\begin{equation}\label{Heq}
B_{\rm eq} = \left(\frac{c_{12}L}{V}\right)^{2/7},
\end{equation}
where $L$ is the radio Luminosity, $V$ is the volume in radiation source, and $c_{12}$ is a numerical constant which is dependent on the spectral index and the upper and lower cut-off frequencies \citet{1970ranp.book.....P}. In \obj, we obtained the spectral indices for S and N components to be $\alpha_{S} \sim -0.6$ and $\alpha_{N} \sim -1$, therefore $c_{12}$ was set as $4.5\times 10^{7}$ and $9.3\times 10^{7}$ for S and N components, respectively. 

The radio luminosity $L$ is an integral of the specific luminosity over the upper and lower cut-off frequencies (100 GHz and 10 MHz, respectively). 
\begin{equation}\label{L}
L=\frac{4 \pi D^{2}_{L}}{(1+z)^{1+\alpha}} \int_{\nu_{1}}^{\nu_{2}} S(\nu) d\nu ,
\end{equation}
in which $D_{L}$ is the luminosity distance.
The volume $V$ is calculated by assuming a spherical geometry, which has a radius of $d_{\rm comp}$. We adopted the fitted full-width-at-half-maximum size of the circular Gaussian model as an estimate of $d_{\rm comp}$ in Table \ref{tab:parm}. 
\begin{equation}\label{V}
V = \frac{4\pi}{3}\left(\frac{D_{L}}{(1+z)^{2}} d_{\rm comp }\right)^{3}
\end{equation}
The derived equipartition magnetic filed strength are tabulated in Table \ref{tab:Bfield}. 
Under the assumption of equipartition between the energy density of the cosmic microwave background and the magnetic field energy density, the magnetic field strength is related to the redshift and can be estimated as: $B_{\rm CMB} = (8\pi/U_{\rm CMB})^{1/2} = 3.26 \times 10^{-6}(1+z)^2$~G \citep{2014MNRAS.438.2694G}.
At $z = 4.56$, we have $B_{\rm CMB} = 0.1$~mG, 2--3 orders of magnitude lower than the calculated equipartition magnetic field strength $B_{\rm eq}$ in the hotspot. This suggests that synchrotron radiation dominates the radiative lose in compact radio sources. In contrast, in extended jets of a few tens of kpc, $B_{\rm eq}$ decreases with the distance from the central engine ($\propto r^{-a}$, $a$ is the power-law index) and may be lower than $B_{\rm CMB}$ starting from some distance, leading to that inverse Compton scattering becomes dominant.

\begin{table}
    \centering
    \caption{Physical parameters of the source components.} 
    \begin{tabular}{ccccc} \hline\hline
    Epoch     & Comp.         & L                     & V                    & $B_{\rm eq}$       \\
    YYYY-MM   &               & ($\rm erg\ s^{-1}$)     & ($\rm cm^{3}$)         & (mG)             \\
    (1)       & (2)           & (3)                   & (4)                  & (5)            \\
    \hline 
    1996-05   & S             & $2.4\times10^{46}$    & $7.5\times10^{57}$   &80               \\
              & N             & $8.4\times10^{45}$    & $1.6\times10^{58}$   &58               \\
    2014-08   & S             & $2.3\times10^{46}$    & $7.5\times10^{57}$   &79               \\
              & N             & $7.5\times10^{45}$    & $2.5\times10^{58}$   &50               \\
    2017-09   & S             & $2.0\times10^{46}$    & $5.7\times10^{57}$   &82               \\
              & N             & $7.2\times10^{45}$    & $6.0\times10^{58}$   &38               \\              
    2018-03   & S             & $2.0\times10^{46}$    & $4.0\times10^{57}$   &91               \\
              & N             & $7.0\times10^{45}$    & $1.7\times10^{58}$   &55               \\    
    \hline 
    \end{tabular} \\ 
    Notes: Col. 1 -- observing date; Col. 2 -- label of the fitted model component; Col. 3 --  radio luminosity; Col. 4 -- volume; Col. 5 -- equipartition magnetic field strength.
    \label{tab:Bfield}
\end{table}

\section{Jet speed and viewing angle}

\begin{figure}
    \centering
    \includegraphics[width=0.45\textwidth]{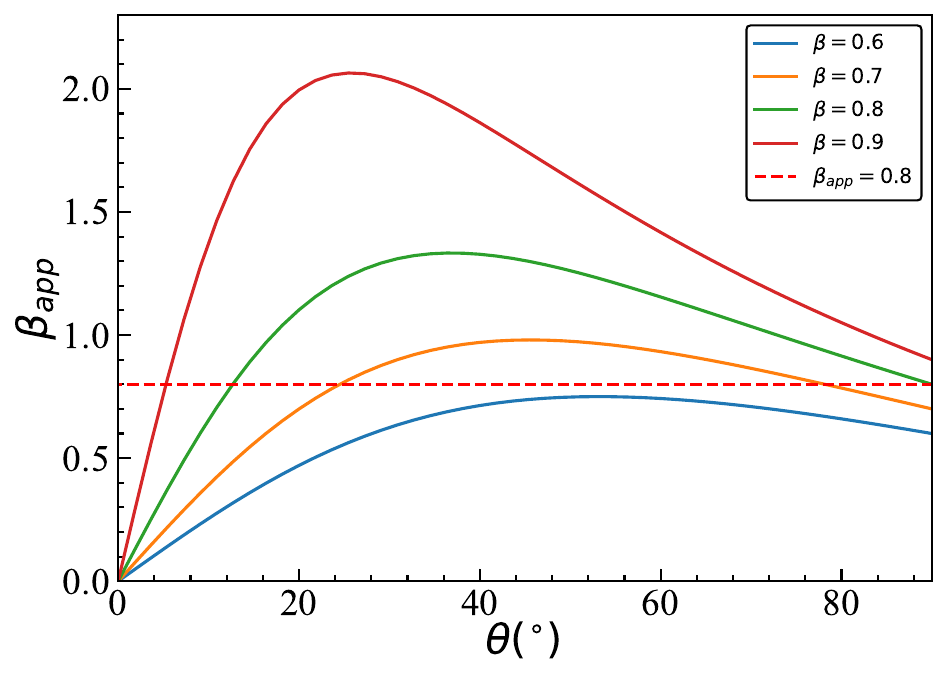}
    \caption{Change of the apparent jet speed $\beta_{\rm app}$ with the jet viewing angle (the angle between the jet direction and the line of sight, $\theta$) for a set of jet speeds $\beta$. }
    \label{fig:app_v}
\end{figure}

Figure \ref{fig:app_v} displays the change of the apparent jet speed ($\beta_{\rm app}$) with the jet viewing angle ($\theta$) for a set of jet speeds ($\beta$). The relation is $\beta_{\rm app} = \frac{\beta\sin\theta}{1 - \beta\cos\theta}$, in which both $\beta_{\rm app}$ and $\beta$ are in unit of the speed of light ($c$).

\bsp	
\label{lastpage}
\end{document}